\documentclass[twocolumn,prd,superscriptaddress,altaffilletter,nofootinbib]{revtex4-1}
\usepackage{amsmath}
\usepackage{amssymb}
\usepackage{amsfonts}
\usepackage{comment}
\usepackage{graphicx}
\usepackage{hyperref}
\usepackage[nolist]{acronym}
\usepackage[usenames,dvipsnames]{xcolor}
\usepackage{xspace}
\usepackage{tikz}
\usepackage{ulem}
\usepackage{siunitx}
\usepackage{multirow}
\usepackage{bm}
\usepackage{romannum}
\usepackage{mathtools}

\usetikzlibrary{arrows,shapes,trees,decorations.pathreplacing}
\DeclareSIUnit \parsec {pc}

\allowdisplaybreaks

\newcommand{\tabref}[1]{Tab. \ref{#1}}
\newcommand{\figref}[1]{Fig.~\ref{#1}}
\newcommand{\secref}[1]{Sec.~\ref{#1}}

\newcommand{\RNum}[1]{\uppercase\expandafter{\romannumeral #1\relax}}

\newcommand{\mA}{m_A}
\newcommand{\eD}{\epsilon_D}
\newcommand{\QD}{Q_D}
\newcommand{\vDM}{v_{\mathrm{DM}}}
\newcommand{\rhoDM}{\rho_{\mathrm{DM}}}
\newcommand{\mn}{m_{\mathrm{n}}}
\newcommand{\flaser}{\nu}
\newcommand{\Teff}{T_{\mathrm{eff}}}
\newcommand{\Tobs}{T_{\mathrm{obs}}}

\newcommand{\Tround}{T_{\mathrm{r}}}
\newcommand{\xin}{x_{\mathrm{i}}}
\newcommand{\xend}{x_{\mathrm{e}}}

\begin{document}

\title{Improved sensitivity of interferometric gravitational wave detectors to ultralight vector dark matter from the finite light-traveling time}

\author{Soichiro Morisaki}
  \affiliation{Department of Physics, University of Wisconsin-Milwaukee, Milwaukee, WI 53201, USA}
\author{Tomohiro Fujita}
  \affiliation{Institute for Cosmic Ray Research, University of Tokyo, Kashiwa, Chiba 277-8582, Japan}
  \author{Yuta Michimura}
  \affiliation{Department of Physics, University of Tokyo, Bunkyo, Tokyo 113-0033, Japan}
  \author{Hiromasa Nakatsuka}
  \affiliation{Institute for Cosmic Ray Research, University of Tokyo, Kashiwa, Chiba 277-8582, Japan}
\author{Ippei Obata}
  \affiliation{Max-Planck-Institut f{\"u}r Astrophysik, Karl-Schwarzschild-Str. 1, 85741 Garching, Germany}
\date{\today}

\begin{abstract}

Recently several studies have pointed out that gravitational-wave detectors are sensitive to ultralight vector dark matter and can improve the current best constraints given by the Equivalence Principle tests. 
While a gravitational-wave detector is a highly precise measuring tool of the length difference of its arms, its sensitivity is limited because the displacements of its test mass mirrors caused by vector dark matter are almost common. 
In this paper we point out that the sensitivity is significantly improved if the effect of finite light-traveling time in the detector's arms is taken into account.
This effect enables advanced LIGO to improve the constraints on the $U(1)_{B-L}$ gauge coupling by an order of magnitude compared with the current best constraints.
It also makes the sensitivities of the future gravitational-wave detectors overwhelmingly better than the current ones.
The factor by which the constraints are improved due to the new effect depends on the mass of the vector dark matter, and the maximum improvement factors are $470$, $880$, $1600$, $180$ and $1400$ for advanced LIGO, Einstein Telescope, Cosmic Explorer, DECIGO and LISA respectively.
Including the new effect, we update the constraints given by the first observing run of advanced LIGO and improve the constraints on the $U(1)_B$ gauge coupling by an order of magnitude compared with the current best constraints.

\end{abstract}

\maketitle

\section{Introduction} \label{introduction}

While the existence of dark mater has been firmly established by the observations, its identity is still unknown.
Weakly Interacting Massive Particles are promising candidates of dark matter, and most of the searches have focused on the electro-weak mass scale \cite{Aprile:2018dbl, Ackermann:2015zua, Sirunyan:2017hci, Aaboud:2016tnv}.
However, despite the extensive efforts, they have not been detected, which motivates us to search for dark matter candidates in different mass range.

Among them is an ultralight boson, whose mass can be down to $\sim10^{-22}~\si{eV}$ \cite{Hu:2000ke}.
Due to the large occupation number, it behaves as classical waves in our Galaxy, whose angular frequency is almost equal to its mass.
A lot of searches have been proposed and conducted to detect this type of dark matter \cite{Khmelnitsky:2013lxt, Porayko:2014rfa, Porayko:2018sfa, Nomura:2019cvc, Arvanitaki:2014faa, Arvanitaki:2015iga, Branca:2016rez, Graham:2015ifn, Blas:2016ddr, Arvanitaki:2016fyj, Geraci:2018fax, Hees:2016gop, Stadnik:2015xbn, Aoki:2016kwl, Morisaki:2018htj, DeRocco:2018jwe, Obata:2018vvr, Liu:2018icu, Nagano:2019rbw, Martynov:2019azm, Michimura:2020vxn, Pierce:2018xmy, Guo:2019ker, Grote:2019uvn, CalderonBustillo:2020srq}. 
Some of them search for the oscillation of fundamental constants such as the fine-structure constant, which may be caused through its coupling to the Standard Model particles \cite{Arvanitaki:2014faa, Arvanitaki:2015iga, Branca:2016rez, Hees:2016gop, Stadnik:2015xbn}.
The metric perturbations generated by it can be detected in the pulsar timing array experiments \cite{Khmelnitsky:2013lxt, Porayko:2014rfa, Porayko:2018sfa, Nomura:2019cvc}.
If it has the axion-type coupling, it differentiates the phase velocities of the circular-polarized photons and may be detected with an optical cavity \cite{DeRocco:2018jwe, Obata:2018vvr, Liu:2018icu, Nagano:2019rbw, Martynov:2019azm} or astronomical observations \cite{Fujita:2018zaj, Caputo:2019tms, Fedderke:2019ajk}.

Recently, it was pointed out that gravitational-wave detectors are sensitive to ultralight vector dark matter arising as a gauge boson of $U(1)_B$ or $U(1)_{B-L}$ gauge symmetry \cite{Pierce:2018xmy}, where $B$ and $L$ are the baryon and lepton numbers, respectively.
The vector dark matter oscillates the test mass mirrors of the detectors though its coupling with baryons or leptons.
Since the gravitational-wave detectors are highly precise measuring tools of the length difference of their arms, they are sensitive to the tiny oscillations, and they can be used to probe the parameter space which has not been excluded by the Equivalence Principle (EP) tests \cite{Schlamminger:2007ht, Wagner:2012ui, Touboul:2017grn, Berge:2017ovy}.
The actual search was also conducted with the data from the first observing run (O1) of the LIGO detectors \cite{Harry:2010zz}, and the constraints better than that from the E\"ot-Wash torsion pendulum experiment \cite{Schlamminger:2007ht, Wagner:2012ui} was obtained for the $U(1)_B$ case \cite{Guo:2019ker}.

What limits the sensitivity of gravitational-wave detectors is that the displacements of the test mass mirrors caused by the vector dark matter are almost common.
It makes the length between the mirrors almost constant over the time, and the amplitude of the signal due to the length change is suppressed by a factor of the velocity of dark matter, which is in the order of $10^{-3}$.
In this paper we point out that the effect of the finite light-traveling time is crucial in this case.
Even if the displacements are completely common, the optical path length of the laser light changes, as the test mass mirrors oscillate while the light is traveling in the arm.
While it is suppressed by the product of oscillation frequency and the arm length, it can be more important than the contribution from the length change.
It becomes more pronounced for the future gravitational-wave detectors, which have longer arms.
This effect was taken into account in the previous studies for scalar dark matter \cite{Arvanitaki:2016fyj, Morisaki:2018htj} but never done before for vector dark matter.

This paper is organized as follows.
In \secref{sec:basics} we introduce the model we consider and the force exerted by the vector dark matter.
In \secref{sec:signal} we calculate the signal produced by the vector dark matter in a gravitational-wave detector taking into account the finite light-traveling time.
In \secref{sec:future} we estimate the future constraints and show how much they are improved due to the new contribution.
In \secref{sec:o1} we update the current constraints from the O1 data of the advanced LIGO detectors. 
Finally we summarize the results we have obtained in \secref{sec:conclusion}.
Throughout this paper we apply the natural unit system, $\hbar=c=\epsilon_0=1$.

\section{Vector dark matter} \label{sec:basics}

We consider a massive vector field, $A^\mu$, which couples to $B$ or $B-L$ current $J^\mu_D~(D=B~\text{or}~B-L)$, as dark matter.
The Lagrangian is given by
\begin{equation}
\mathcal{L} = -\frac{1}{4} F^{\mu \nu} F_{\mu \nu} + \frac{1}{2} \mA^2 A^\mu A_\mu - \eD e J^\mu_D A_\mu, \label{lagrangian}
\end{equation}
where $F_{\mu \nu} = \partial_\mu A_\nu - \partial_\nu A_\mu$, $\mA$ is the mass of the vector field and $\eD$ is the coupling constant normalized to the electromagnetic one $e$.  

The spatial components of the vector dark matter in our Galaxy can be modeled as \cite{Miller:2020vsl}
\begin{equation}
\bm{A} = \sum_i A_i \bm{e}_i \cos (\omega_i t - \bm{k}_i \cdot \bm{x} + \phi_i),
\end{equation}
where $i$ is an index to identify each dark matter particle and we sum over their vector potentials.
$A_i$ is the amplitude, $\bm{e}_i$ is the polarization unit vector, $\omega_i$ is the angular frequency, $\bm{k}_i$ is the wave number and $\phi_i$ is the constant phase of the $i$-th particle.
The equation of motion gives the following dispersion relation,
\begin{equation}
\omega_i = \sqrt{\bm{k}^2_i + \mA^2}. \label{eq:dispersion}
\end{equation}

The norms of the wave numbers in our Galaxy are in the order of $\mA \vDM \sim 10^{-3} \mA$, where $\vDM$ is the dark matter velocity dispersion in our Galaxy.
Substituting it into \eqref{eq:dispersion} leads to
\begin{equation}
\omega_i - \mA \sim \mA \vDM^2 \sim 10^{-6} \mA.
\end{equation}
This means the vector field, and hence the signal we observe, can be treated as monochromatic waves with frequency of $\mA / 2 \pi$ over the coherence time, which is given by
\begin{equation}
\tau \equiv \frac{2 \pi}{\mA \vDM^2} \sim \frac{10^7}{\mA},
\end{equation}
and the coherence is lost for a longer time interval.

The force exerted by the vector dark matter on a test mass mirror located at $\bm{x}_0$ is given by
\begin{align}
\bm{F} &\simeq - \eD e \QD \dot{\bm{A}} \nonumber \\
&\simeq \mA \eD e \QD \sum_i A_i \bm{e}_i \sin(\omega_i t - \bm{k}_i \cdot \bm{x}_0 + \phi_i),
\end{align}
where $\QD$ is the $B$ or $B-L$ charge of the test mass mirror.
The test mass mirror oscillates around $\bm{x}_0$ due to the force, and its position is given by $\bm{x}=\bm{x}_0 + \delta \bm{x}(t, \bm{x}_0)$, where
\begin{equation}
\delta \bm{x}(t, \bm{x}_0) \simeq - \frac{\eD e}{\mA} \frac{\QD}{M} \sum_i A_i \bm{e}_i \sin(\omega_i t - \bm{k}_i \cdot \bm{x}_0 + \phi_i).
\end{equation}
$\QD/M$ is approximately given by
\begin{equation}
\frac{\QD}{M} \simeq \begin{dcases}
\displaystyle \frac{1}{\mn}, & \left(D=B\right) \\
\displaystyle \frac{0.5}{\mn}, & \left(D=B-L\right) 
\end{dcases}
\end{equation}
where $\mn$ is the neutron mass.

\section{Signal in a gravitational-wave detector} \label{sec:signal}

The signal in a gravitational-wave detector is given by
\begin{equation}
h(t)=\frac{\varphi(t; \bm{n}) - \varphi(t; \bm{m})}{4 \pi \flaser L}, \label{eq:hoft}
\end{equation}
where $\flaser$ is the laser frequency of the detector, $L$ is the arm length, and $\bm{n}$ and $\bm{m}$ are unit vectors along the two arms of the interferometer.
$\varphi(t; \bm{n})$ is the phase of laser light returning back from the arm after the round trip.

The phase of laser light returning back at the time $t$ is the same as that of laser light entering the arm at the time $t-\Tround$, where $\Tround$ is the round-trip time.
Thus, we have
\begin{equation}
\varphi(t; \bm{n}) = 2 \pi \flaser (t - \Tround) + \phi_0, \label{eq:returnphase}
\end{equation}
where $\phi_0$ is a constant phase.
The round-trip time is given by
\begin{equation}
\Tround = -\xin (t) + 2 \xend(t-L) - \xin(t-2L), \label{eq:roundtime}
\end{equation}
where $\xin(t)$ and $\xend(t)$ represent the positions of the input and end test mass mirrors of the arm.
With the coordinate system where the input test mass mirror is at $\bm{x}=0$ in the absence of vector dark matter, $\xin(t)$ and $\xend(t)$ are given by
\begin{equation}
\xin(t) = \bm{n} \cdot \delta \bm{x}(t, \bm{0}),~~~~~\xend(t) = L + \bm{n} \cdot \delta \bm{x}(t, L \bm{n}). \label{eq:xin_xend}
\end{equation}
Substituting \eqref{eq:roundtime} and \eqref{eq:xin_xend} into \eqref{eq:returnphase}, we obtain
\begin{equation}
\varphi(t; \bm{n}) = - 2 \pi \nu \left(\delta L_1 + \delta L_2\right) + 2 \pi \flaser (t - 2 L) + \phi_0,
\end{equation}
where
\begin{align}
\delta L_1 &\equiv \bm{n} \cdot \left(-\delta \bm{x}(t, \bm{0}) + 2 \delta \bm{x}(t - L, \bm{0}) - \delta \bm{x} (t - 2 L, \bm{0})\right) \nonumber \\
&= - \frac{4 \eD e}{\mA} \frac{\QD}{M} \sin^2\left(\frac{\mA L}{2}\right) \nonumber \\
&~~\times \sum_i A_i \left(\bm{n} \cdot \bm{e}_i\right) \sin\left(\omega_i (t - L) + \phi_i\right), \\
\delta L_2 &\equiv 2 \bm{n} \cdot \left( \delta \bm{x}(t - L, L \bm{n}) - \delta \bm{x}(t - L, \bm{0}) \right) \nonumber \\
&\simeq \frac{2 \eD e L}{\mA} \frac{\QD}{M} \nonumber \\
&~~ \times \sum_i A_i \left(\bm{n}\cdot\bm{e}_i\right) \left(\bm{n}\cdot\bm{k}_i\right) \cos \left(\omega_i (t-L) + \phi_i\right).
\end{align}
To derive the approximate expression of $\delta L_2$, we assume $L| \bm{k}|\ll1$, which is valid for the frequency range and the arm length of the gravitational-wave detectors we consider.

As can be seen in the definition of $\delta L_2$, it is from the deviation of the arm length from $L$, and it has been taken into account in the previous studies.
Compared to the gravitational waves with the same frequency, the wavelength of the vector dark matter is longer by a factor of $1/\vDM \sim 10^3$.
This makes force acting on the two test mass mirrors at both ends of the arm almost the same, and $\delta L_2$ is suppressed by a factor of $\vDM \sim 10^{-3}$ through $\bm{k}_i$.

On the other hand, $\delta L_1$ is the new contribution we point out, which arises due to the finite light-traveling time in the arm.
Even if the displacements are completely common and the arm length is constant, the optical path length can oscillate, as the test mass mirrors oscillate while light is traveling.
This contribution is significant only when the oscillation frequency is comparable to the inverse of the round-trip time, and it is suppressed by $\sim \mA L$.
Nevertheless, $\delta L_1$ is important in this case as $\delta L_2$ is suppressed more significantly.
For the advanced LIGO detector, whose arm length is $4~\si{\kilo \metre}$ and frequency band is $10-1000~\si{\hertz}$, the ratio between $\delta L_1$ and $\delta L_2$ is given by
\begin{equation}
\frac{\delta L_1}{\delta L_2} \sim \frac{\mA L}{\vDM} \sim 8 \left( \frac{\mA}{2 \pi \times 100\si{\hertz}} \right),
\end{equation}
which indicates $\delta L_1$ is more significant in most of the frequency range.
The ratio becomes larger for the future detectors, whose arms are longer, and the improvements due to $\delta L_1$ are more pronounced as shown in the next section.

$\varphi(t; \bm{m})$ can be calculated just by replacing $\bm{n}$ by $\bm{m}$ in $\varphi(t; \bm{n})$, and the signal is given by
\begin{equation}
h(t)= h_1(t) + h_2(t),
\end{equation}
where
\begin{align}
&h_1(t) = \frac{2 \eD e}{\mA L} \frac{\QD}{M} \sin^2\left(\frac{\mA L}{2}\right) \nonumber \\
&~~\times \sum_i A_i \left(\bm{n} \cdot \bm{e}_i - \bm{m} \cdot \bm{e}_i \right) \sin\left(\omega_i (t - L) + \phi_i\right),  \\ 
&h_2(t) = - \frac{\eD e}{\mA} \frac{\QD}{M} \sum_i A_i \big( \left(\bm{n}\cdot\bm{e}_i\right) \left(\bm{n}\cdot\bm{k}_i\right) \nonumber \\
&~~-  \left(\bm{m}\cdot\bm{e}_i\right) \left(\bm{m}\cdot\bm{k}_i\right)\big) \cos \left(\omega_i (t-L) + \phi_i\right).
\end{align}
$h_1$ and $h_2$ come from $\delta L_1$ and $\delta L_2$ respectively, and $h_1$ is the new contribution to the signal.
Most of the detectors we consider form Fabry-P\'{e}rot cavities, which amplify the signal.
However, the amplification factors are taken into account in the sensitivity curves, and we do not need to consider them in calculating the signal.

\section{Future prospects} \label{sec:future}

We estimate the sensitivities achieved by the future gravitational-wave experiments, taking into account the new contribution $h_1$.
Here we consider advanced LIGO (aLIGO), Einstein telescope (ET) \cite{Hild:2010id}, Cosmic Explorer (CE) \cite{Evans:2016mbw}, DECIGO \cite{Kawamura:2006up} and LISA \cite{Audley:2017drz} as representative gravitational-wave detectors.

The signal keeps its coherence only for the finite time of $\tau$.
One of the detection methods suitable for this type of signal is the semi-coherent method \cite{Morisaki:2018htj, Miller:2020vsl}, where the whole data are split into segments whose lengths are $\sim \tau$ and the squares of the Fourier components calculated with the segments are summed up incoherently.
The detection threshold of the signal's amplitude with this detection method can be estimated with
\begin{equation}
\left<h^2\right> = \frac{S\left(\frac{\mA}{2\pi}\right)}{\Teff}. \label{threshold}
\end{equation}
While the previous study \cite{Pierce:2018xmy} considered a different detection method, which correlates data from multiple detectors, the difference of the threshold amplitude is within an $\mathcal{O}(1)$ factor \cite{Morisaki:2018htj}.

\begin{figure}
\includegraphics[width=0.95\hsize]{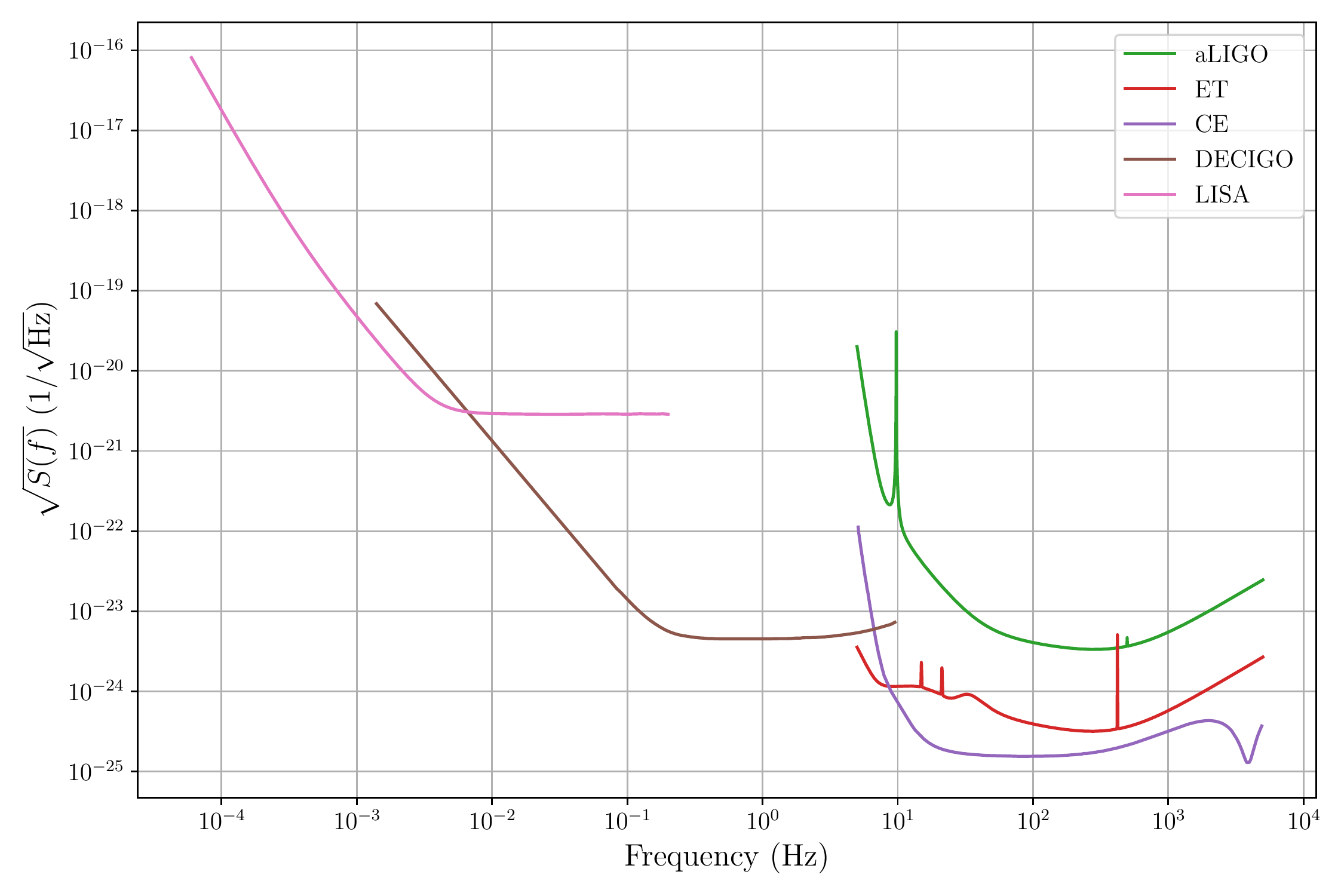}
  \caption{The one-sided power spectra of noise in the $h(t)$ channel given by \eqref{eq:hoft}. Those for advanced LIGO (aLIGO), Einstein Telescope (ET), Cosmic Explorer (CE) and DECIGO are taken from \cite{aligosensitivity}, \cite{Evans:2016mbw}, \cite{Essick:2017wyl} and \cite{Kawamura:2020pcg} respectively. The noise spectrum for LISA is calculated with the target acceleration and displacement noise level given in \cite{Audley:2017drz}.}
  \label{sensitivity}
\end{figure}

$S(f)$ is the one-sided power spectral density (PSD) of noise in the $h(t)$ channel.
The PSDs for the representative detectors are shown in \figref{sensitivity}.
$\Teff$ is the effective observation time given by
\begin{equation}
\Teff = \begin{dcases}
\displaystyle \Tobs, & \left(\Tobs < \tau \right) \\
\displaystyle \sqrt{\tau \Tobs}, & \left(\Tobs \geq \tau \right) \
\end{dcases}
\end{equation}
where $\Tobs$ is the observational time.
$\left<h^2\right>$ is $h^2(t)$ averaged over time.
Averaging over random polarization and propagation directions, we can estimate it as follows,
\begin{align}
&\left< h^2 \right> = \left< h^2_1 \right> + \left< h^2_2 \right>, \label{hsquare} \\
&\left< h^2_1 \right> = \frac{8 \eD^2 e^2 \rhoDM}{3 \mA^4 L^2} \frac{\QD^2}{M^2} \sin^4 \left(\frac{\mA L}{2}\right) \left(1 - \bm{n} \cdot \bm{m}\right), \label{hsquare1} \\
&\left< h^2_2 \right> = \frac{2 \eD^2 e^2 \vDM^2 \rhoDM}{9 \mA^2} \frac{\QD^2}{M^2} \left(1 - \left(\bm{n}\cdot\bm{m}\right)^2\right). \label{hsquare2}
\end{align}
The values of the arm length, $L$, for the representative detectors are listed in \tabref{armlength}.
$\bm{n} \cdot \bm{m}=0$ for aLIGO and CE, and $\bm{n} \cdot \bm{m}= 1/2$ for ET, DECIGO and LISA.

\begin{figure}
\includegraphics[width=0.95\hsize]{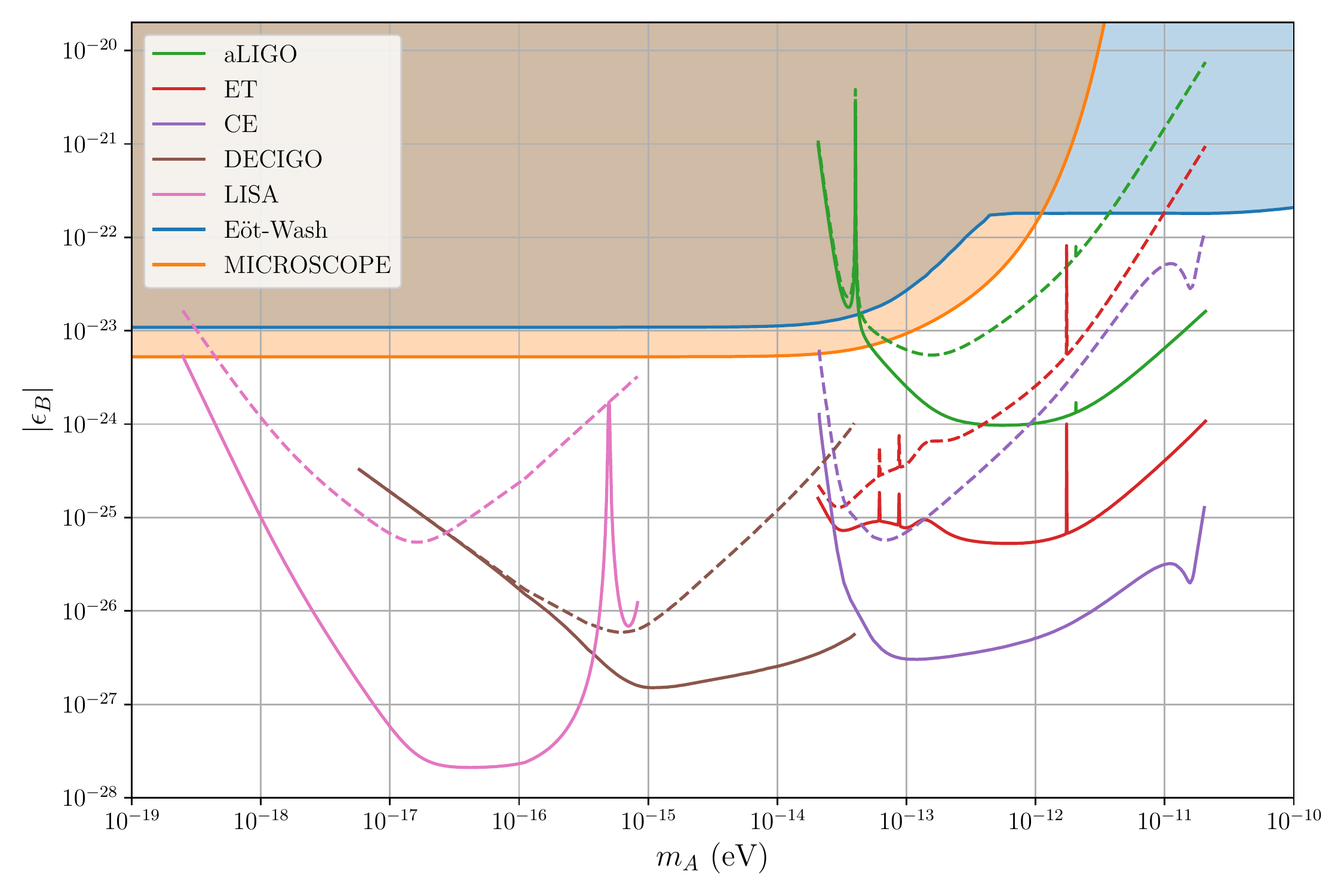}
\includegraphics[width=0.95\hsize]{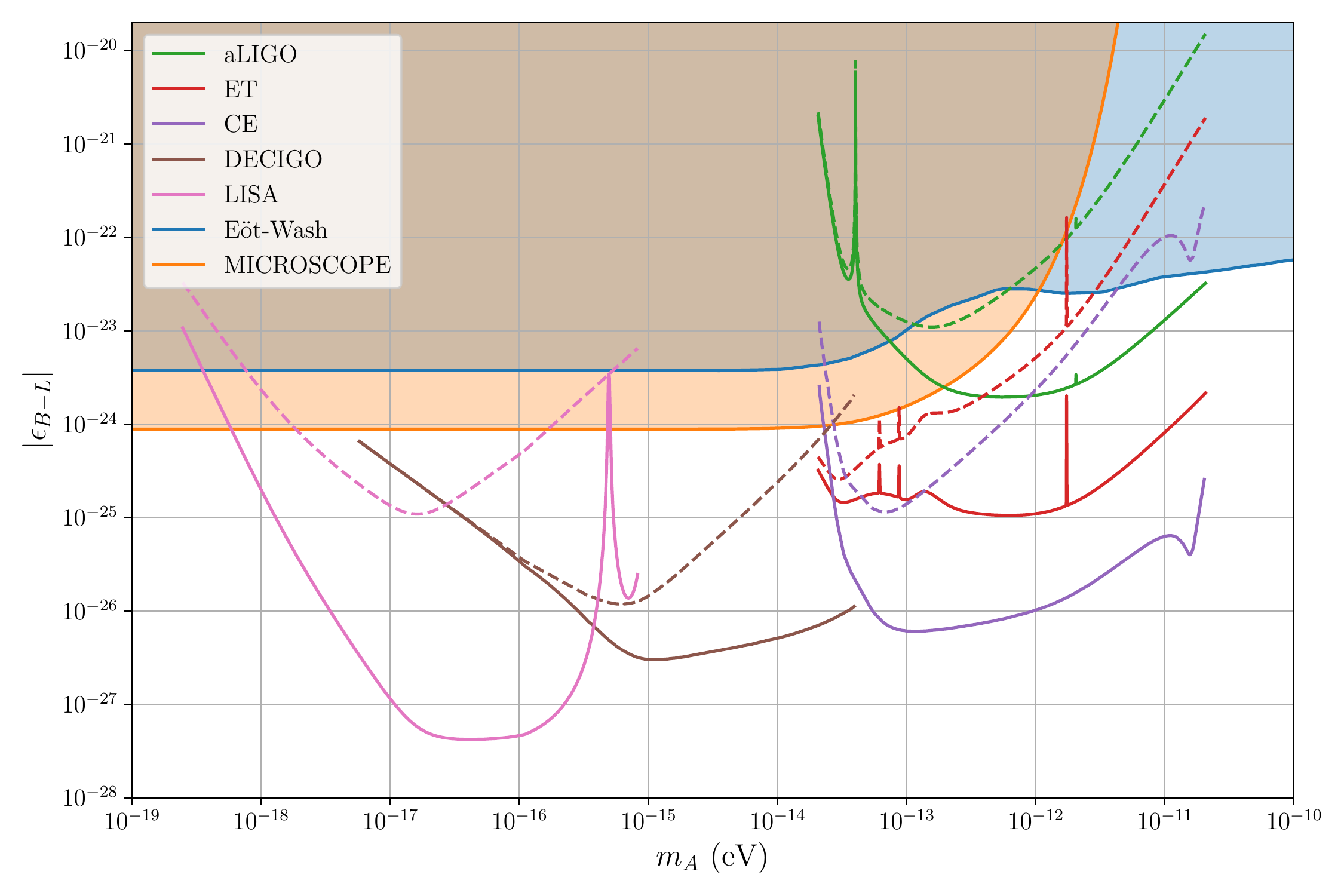}
\caption{The constraints on the coupling constant given by the future gravitational-wave detectors. We assume the observation time of $2$ years. For comparison, the constraints given by the E\"ot-Wash experiment \cite{Schlamminger:2007ht, Wagner:2012ui} and the MICROSCOPE experiment \cite{Touboul:2017grn, Berge:2017ovy, Fayet:2017pdp}, which are the tests of the Equivalence Principle, are also shown. The shaded region has been already excluded. The dashed lines represent the constraints if the effects of the finite light-traveling time were not present. The upper figure shows the constraints for $U(1)_B$ and the lower one for $U(1)_{B-L}$.}
  \label{future}
\end{figure}

The future constraints on $|\eD|$ estimated with \eqref{threshold} are shown in \figref{future}.
Here, we assume the observation time of $2$ years and apply $\vDM=230~\si{\kilo \metre.\second^{-1}}$, which is taken from \cite{Smith:2006ym} and applied in \cite{Pierce:2018xmy}.
For comparison, the constraints without the contribution from $h_1$ are shown as dashed lines.
The figure shows that the inclusion of $h_1$ significantly improves the constraints.
The factor by which the constraints are improved depends on the mass, and the maximum improvement factors are $470$, $880$, $1600$, $180$ and $1400$ for aLIGO, ET, CE, DECIGO and LISA respectively.
The improvements are more significant for ET and CE compared to aLIGO because they have longer arms.
The relatively significant improvement for LISA is due to its long arm length.
The contribution from $h_1$ is canceled at $\mA\simeq5\times10^{-16}~\si{\eV}$ for LISA, where the frequency of the signal is equal to the inverse of the one-way-trip time of the light.

While the constraints without the contribution from $h_1$ should correspond to those calculated in \cite{Pierce:2018xmy}, our constraints of LISA are significantly better than their $2$-$\sigma$ constraints.
After the error of a factor of 2 in their constraints on $\eD$, which was pointed out in \cite{Guo:2019ker}, is corrected, our constraints are better than their constraints by an order of magnitude at $\mA \gtrsim 10^{-16}~\si{eV}$.
The reasons for the difference are as follows:
(1) Our order-of-estimate constraints correspond to the $1$-$\sigma$ constraints in their analysis.
(2) They considered correlating the data from multiple channels, and the signal-to-noise ratio is degraded by a factor of the overlap reduction function \cite{Allen:1997ad}.
(3) The PSD they used is for gravitational-wave strain and increases in proportion to $f^2$ at high frequency due to the cancellation of the signal at $f \gtrsim (2 L)^{-1}$ \cite{Larson:1999we}. Such signal cancellation does not occur for $h_2$ of dark matter signal as its wavelength is much longer than that of gravitational waves with the same frequency.
(4) They averaged the amplitude of the signal over the directions of vector field and its momentum while they used the PSD averaged over the polarization angle and propagation direction of gravitational waves, which resulted in double counting of the geometric factor.

The current best constraints given by the EP tests are also shown as blue and orange lines in \figref{future}.
The figure shows that the $h_1$ contribution makes the future constraints better than the current best constraints by orders of magnitude for both $U(1)_B$ and $U(1)_{B-L}$ cases.
For reference, the constraints are improved by factors of $23000$, $2100$ and $27000$ at $\mA=10^{-16}~\si{\eV}$, $10^{-14}~\si{\eV}$ and $10^{-12}~\si{\eV}$ respectively for the $U(1)_B$ case, and $1900$, $180$ and $2200$ respectively for the $U(1)_{B-L}$ case.
Notably the inclusion of $h_1$ enables aLIGO to improve the constraints on the $U(1)_{B-L}$ gauge coupling by an order of magnitude.

\begin{table}
    \caption{The arm length of the gravitational-wave detectors. All the values are in unit of $\si{\metre}$.}
\begin{ruledtabular}
\begin{tabular}{ccccc}
 aLIGO & ET & CE & DECIGO & LISA \\
\hline
$4\times10^3$ & $1\times10^4$ & $4 \times 10^4$ & $1 \times 10^6$ & $2.5 \times 10^9$ \\
\end{tabular}
\end{ruledtabular}
\label{armlength}
\end{table}

\section{Advanced LIGO O1} \label{sec:o1}

\begin{figure}
\includegraphics[width=0.95\hsize]{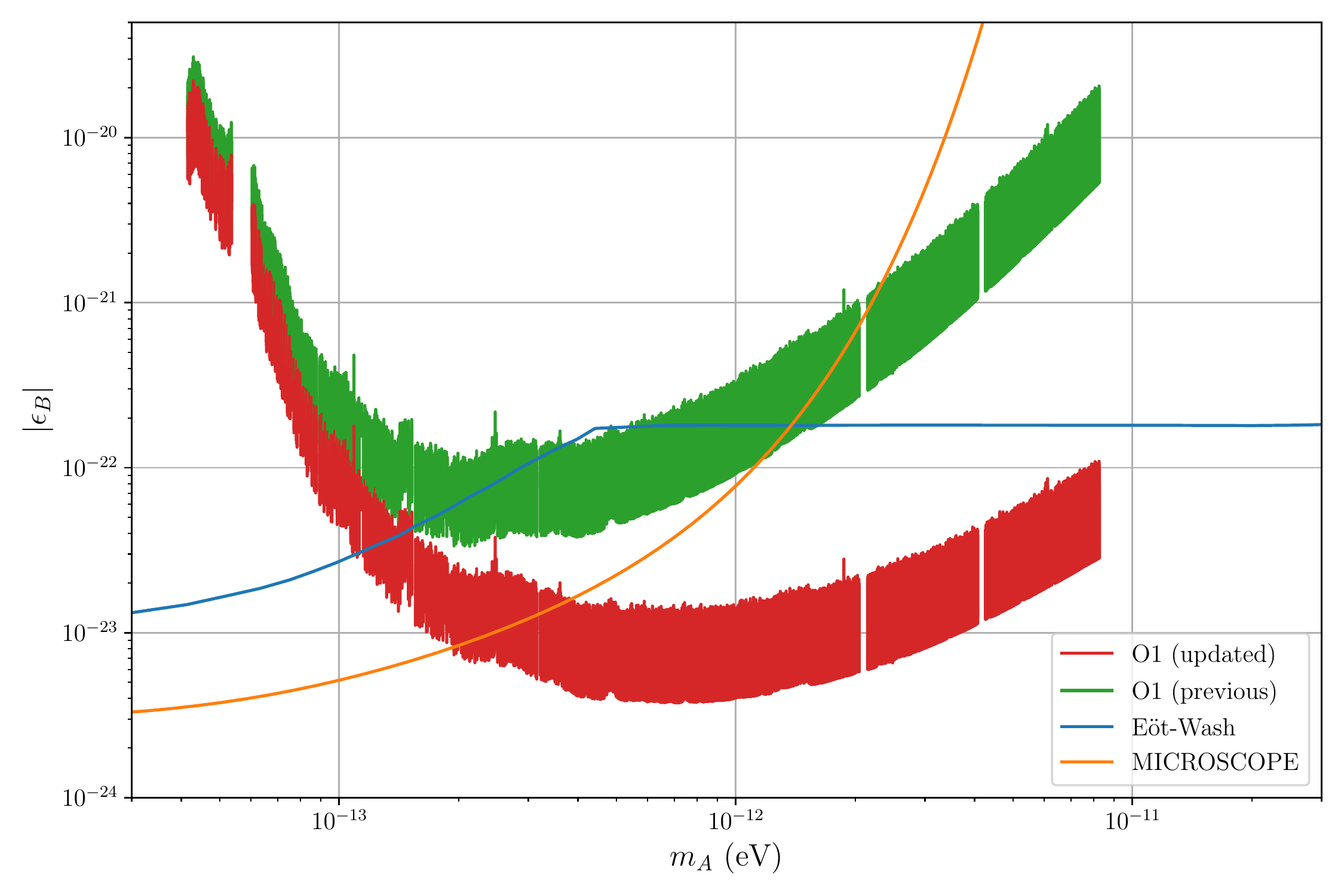}
\includegraphics[width=0.95\hsize]{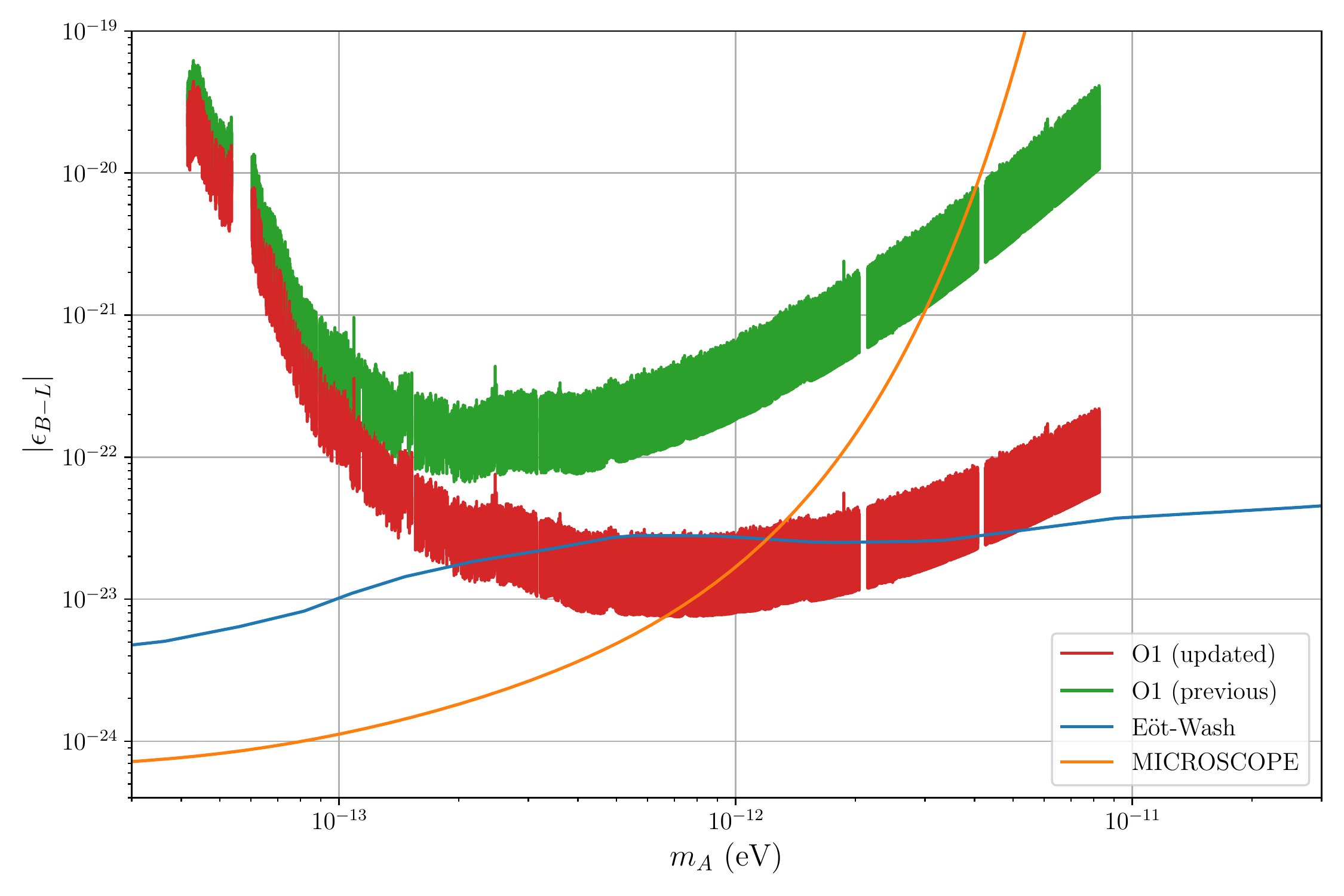}
  \caption{The constraints given by the LIGO O1, which are updated by the inclusion of the effect from the finite light-traveling time, are shown in red. The green lines represent the constraints previously calculated without the effect. The orange and blue lines represent the constraints from the Equivalence Principle tests. The upper figure is for the $U(1)_B$ case and the lower one is for the $U(1)_{B-L}$ case.}
  \label{O1}
\end{figure}

We update the constraints given by the O1 data of aLIGO by incorporating $h_1$.
The inclusion of $h_1$ improves the constraints by a factor of $\sqrt{\left(\left<h^2_1\right> + \left<h^2_2\right>\right) / \left<h^2_2\right>}$, and the improved constraints are shown as red lines in \figref{O1}.
The previously calculated constraints are also shown as green lines.
As seen in the figure, the inclusion of $h_1$ makes the O1 constraints on the $U(1)_B$ gauge coupling better than the current best constraints at $\mA\gtrsim2 \times 10^{-13}~\si{\eV}$ and better by an order of magnitude around $\mA = 10^{-12}~\si{\eV}$.
The improved O1 constraints on the $U(1)_{B-L}$ gauge coupling are comparable to the current best constraints at $7 \times 10^{-13}~\si{\eV} \lesssim \mA \lesssim 5 \times 10^{-12}~\si{\eV}$.

\section{Conclusion} \label{sec:conclusion}

In this paper we have pointed out that the effect of the finite light-traveling time is crucial for calculating the signal produced by ultralight vector dark matter in a gravitational-wave detector.
By taking it into account properly we have calculated the new contribution to the signal.
Then we have estimated the future constraints on the gauge coupling given by gravitational-wave detectors incorporating the new contribution.
As a results, we have found that the new contribution significantly improves the future constraints given by gravitational-wave detectors.
The factor by which the constraints are improved depends on the mass of the vector dark matter, and the maximum improvement factors are $470$, $880$, $1600$, $180$ and $1400$ for aLIGO, ET, CE, DECIGO and LISA respectively.
These improvements make the future constraints better than the current best constraints from the EP tests by orders of magnitude.
Notably, it enables aLIGO to improve the constraints on the $U(1)_{B-L}$ gauge coupling by an order of magnitude.

Finally, we have updated the constraints given by the aLIGO O1 data incorporating the new contribution.
The updated constraints on the $U(1)_B$ gauge coupling are better than the current best constraints by an order of magnitude around $\mA = 10^{-12}~\si{\eV}$.

\begin{acknowledgments}

This work was supported by JSPS KAKENHI Grant Numbers 18H01224, 18K13537, 18K18763, 19J21974, 20H05850, 20H05854, 20H05859, and NSF PHY-1912649. H.N. is supported by the Advanced Leading Graduate Course for Photon Science, and I.O. is supported by the JSPS Overseas Research Fellowship.

\end{acknowledgments}

\bibliographystyle{apsrev4-1}
\bibliography{reference}

\begin{thebibliography}{50}%
\makeatletter
\providecommand \@ifxundefined [1]{%
 \@ifx{#1\undefined}
}%
\providecommand \@ifnum [1]{%
 \ifnum #1\expandafter \@firstoftwo
 \else \expandafter \@secondoftwo
 \fi
}%
\providecommand \@ifx [1]{%
 \ifx #1\expandafter \@firstoftwo
 \else \expandafter \@secondoftwo
 \fi
}%
\providecommand \natexlab [1]{#1}%
\providecommand \enquote  [1]{``#1''}%
\providecommand \bibnamefont  [1]{#1}%
\providecommand \bibfnamefont [1]{#1}%
\providecommand \citenamefont [1]{#1}%
\providecommand \href@noop [0]{\@secondoftwo}%
\providecommand \href [0]{\begingroup \@sanitize@url \@href}%
\providecommand \@href[1]{\@@startlink{#1}\@@href}%
\providecommand \@@href[1]{\endgroup#1\@@endlink}%
\providecommand \@sanitize@url [0]{\catcode `\\12\catcode `\$12\catcode
  `\&12\catcode `\#12\catcode `\^12\catcode `\_12\catcode `\%12\relax}%
\providecommand \@@startlink[1]{}%
\providecommand \@@endlink[0]{}%
\providecommand \url  [0]{\begingroup\@sanitize@url \@url }%
\providecommand \@url [1]{\endgroup\@href {#1}{\urlprefix }}%
\providecommand \urlprefix  [0]{URL }%
\providecommand \Eprint [0]{\href }%
\providecommand \doibase [0]{http://dx.doi.org/}%
\providecommand \selectlanguage [0]{\@gobble}%
\providecommand \bibinfo  [0]{\@secondoftwo}%
\providecommand \bibfield  [0]{\@secondoftwo}%
\providecommand \translation [1]{[#1]}%
\providecommand \BibitemOpen [0]{}%
\providecommand \bibitemStop [0]{}%
\providecommand \bibitemNoStop [0]{.\EOS\space}%
\providecommand \EOS [0]{\spacefactor3000\relax}%
\providecommand \BibitemShut  [1]{\csname bibitem#1\endcsname}%
\let\auto@bib@innerbib\@empty
\bibitem [{\citenamefont {Aprile}\ \emph {et~al.}(2018)\citenamefont {Aprile}
  \emph {et~al.}}]{Aprile:2018dbl}%
  \BibitemOpen
  \bibfield  {author} {\bibinfo {author} {\bibfnamefont {E.}~\bibnamefont
  {Aprile}} \emph {et~al.} (\bibinfo {collaboration} {XENON}),\ }\href
  {\doibase 10.1103/PhysRevLett.121.111302} {\bibfield  {journal} {\bibinfo
  {journal} {Phys. Rev. Lett.}\ }\textbf {\bibinfo {volume} {121}},\ \bibinfo
  {pages} {111302} (\bibinfo {year} {2018})},\ \Eprint
  {http://arxiv.org/abs/1805.12562} {arXiv:1805.12562 [astro-ph.CO]}
  \BibitemShut {NoStop}%
\bibitem [{\citenamefont {Ackermann}\ \emph {et~al.}(2015)\citenamefont
  {Ackermann} \emph {et~al.}}]{Ackermann:2015zua}%
  \BibitemOpen
  \bibfield  {author} {\bibinfo {author} {\bibfnamefont {M.}~\bibnamefont
  {Ackermann}} \emph {et~al.} (\bibinfo {collaboration} {Fermi-LAT}),\ }\href
  {\doibase 10.1103/PhysRevLett.115.231301} {\bibfield  {journal} {\bibinfo
  {journal} {Phys. Rev. Lett.}\ }\textbf {\bibinfo {volume} {115}},\ \bibinfo
  {pages} {231301} (\bibinfo {year} {2015})},\ \Eprint
  {http://arxiv.org/abs/1503.02641} {arXiv:1503.02641 [astro-ph.HE]}
  \BibitemShut {NoStop}%
\bibitem [{\citenamefont {Sirunyan}\ \emph {et~al.}(2017)\citenamefont
  {Sirunyan} \emph {et~al.}}]{Sirunyan:2017hci}%
  \BibitemOpen
  \bibfield  {author} {\bibinfo {author} {\bibfnamefont {A.~M.}\ \bibnamefont
  {Sirunyan}} \emph {et~al.} (\bibinfo {collaboration} {CMS}),\ }\href
  {\doibase 10.1007/JHEP07(2017)014} {\bibfield  {journal} {\bibinfo  {journal}
  {JHEP}\ }\textbf {\bibinfo {volume} {07}},\ \bibinfo {pages} {014} (\bibinfo
  {year} {2017})},\ \Eprint {http://arxiv.org/abs/1703.01651} {arXiv:1703.01651
  [hep-ex]} \BibitemShut {NoStop}%
\bibitem [{\citenamefont {Aaboud}\ \emph {et~al.}(2016)\citenamefont {Aaboud}
  \emph {et~al.}}]{Aaboud:2016tnv}%
  \BibitemOpen
  \bibfield  {author} {\bibinfo {author} {\bibfnamefont {M.}~\bibnamefont
  {Aaboud}} \emph {et~al.} (\bibinfo {collaboration} {ATLAS}),\ }\href
  {\doibase 10.1103/PhysRevD.94.032005} {\bibfield  {journal} {\bibinfo
  {journal} {Phys. Rev.}\ }\textbf {\bibinfo {volume} {D94}},\ \bibinfo {pages}
  {032005} (\bibinfo {year} {2016})},\ \Eprint
  {http://arxiv.org/abs/1604.07773} {arXiv:1604.07773 [hep-ex]} \BibitemShut
  {NoStop}%
\bibitem [{\citenamefont {Hu}\ \emph {et~al.}(2000)\citenamefont {Hu},
  \citenamefont {Barkana},\ and\ \citenamefont {Gruzinov}}]{Hu:2000ke}%
  \BibitemOpen
  \bibfield  {author} {\bibinfo {author} {\bibfnamefont {W.}~\bibnamefont
  {Hu}}, \bibinfo {author} {\bibfnamefont {R.}~\bibnamefont {Barkana}}, \ and\
  \bibinfo {author} {\bibfnamefont {A.}~\bibnamefont {Gruzinov}},\ }\href
  {\doibase 10.1103/PhysRevLett.85.1158} {\bibfield  {journal} {\bibinfo
  {journal} {Phys. Rev. Lett.}\ }\textbf {\bibinfo {volume} {85}},\ \bibinfo
  {pages} {1158} (\bibinfo {year} {2000})},\ \Eprint
  {http://arxiv.org/abs/astro-ph/0003365} {arXiv:astro-ph/0003365} \BibitemShut
  {NoStop}%
\bibitem [{\citenamefont {Khmelnitsky}\ and\ \citenamefont
  {Rubakov}(2014)}]{Khmelnitsky:2013lxt}%
  \BibitemOpen
  \bibfield  {author} {\bibinfo {author} {\bibfnamefont {A.}~\bibnamefont
  {Khmelnitsky}}\ and\ \bibinfo {author} {\bibfnamefont {V.}~\bibnamefont
  {Rubakov}},\ }\href {\doibase 10.1088/1475-7516/2014/02/019} {\bibfield
  {journal} {\bibinfo  {journal} {JCAP}\ }\textbf {\bibinfo {volume} {1402}},\
  \bibinfo {pages} {019} (\bibinfo {year} {2014})},\ \Eprint
  {http://arxiv.org/abs/1309.5888} {arXiv:1309.5888 [astro-ph.CO]} \BibitemShut
  {NoStop}%
\bibitem [{\citenamefont {Porayko}\ and\ \citenamefont
  {Postnov}(2014)}]{Porayko:2014rfa}%
  \BibitemOpen
  \bibfield  {author} {\bibinfo {author} {\bibfnamefont {N.~K.}\ \bibnamefont
  {Porayko}}\ and\ \bibinfo {author} {\bibfnamefont {K.~A.}\ \bibnamefont
  {Postnov}},\ }\href {\doibase 10.1103/PhysRevD.90.062008} {\bibfield
  {journal} {\bibinfo  {journal} {Phys. Rev.}\ }\textbf {\bibinfo {volume}
  {D90}},\ \bibinfo {pages} {062008} (\bibinfo {year} {2014})},\ \Eprint
  {http://arxiv.org/abs/1408.4670} {arXiv:1408.4670 [astro-ph.CO]} \BibitemShut
  {NoStop}%
\bibitem [{\citenamefont {Porayko}\ \emph {et~al.}(2018)\citenamefont {Porayko}
  \emph {et~al.}}]{Porayko:2018sfa}%
  \BibitemOpen
  \bibfield  {author} {\bibinfo {author} {\bibfnamefont {N.~K.}\ \bibnamefont
  {Porayko}} \emph {et~al.},\ }\href@noop {} {\  (\bibinfo {year} {2018})},\
  \Eprint {http://arxiv.org/abs/1810.03227} {arXiv:1810.03227 [astro-ph.CO]}
  \BibitemShut {NoStop}%
\bibitem [{\citenamefont {Nomura}\ \emph {et~al.}(2020)\citenamefont {Nomura},
  \citenamefont {Ito},\ and\ \citenamefont {Soda}}]{Nomura:2019cvc}%
  \BibitemOpen
  \bibfield  {author} {\bibinfo {author} {\bibfnamefont {K.}~\bibnamefont
  {Nomura}}, \bibinfo {author} {\bibfnamefont {A.}~\bibnamefont {Ito}}, \ and\
  \bibinfo {author} {\bibfnamefont {J.}~\bibnamefont {Soda}},\ }\href {\doibase
  10.1140/epjc/s10052-020-7990-y} {\bibfield  {journal} {\bibinfo  {journal}
  {Eur. Phys. J. C}\ }\textbf {\bibinfo {volume} {80}},\ \bibinfo {pages} {419}
  (\bibinfo {year} {2020})},\ \Eprint {http://arxiv.org/abs/1912.10210}
  {arXiv:1912.10210 [gr-qc]} \BibitemShut {NoStop}%
\bibitem [{\citenamefont {Arvanitaki}\ \emph {et~al.}(2015)\citenamefont
  {Arvanitaki}, \citenamefont {Huang},\ and\ \citenamefont
  {Van~Tilburg}}]{Arvanitaki:2014faa}%
  \BibitemOpen
  \bibfield  {author} {\bibinfo {author} {\bibfnamefont {A.}~\bibnamefont
  {Arvanitaki}}, \bibinfo {author} {\bibfnamefont {J.}~\bibnamefont {Huang}}, \
  and\ \bibinfo {author} {\bibfnamefont {K.}~\bibnamefont {Van~Tilburg}},\
  }\href {\doibase 10.1103/PhysRevD.91.015015} {\bibfield  {journal} {\bibinfo
  {journal} {Phys. Rev.}\ }\textbf {\bibinfo {volume} {D91}},\ \bibinfo {pages}
  {015015} (\bibinfo {year} {2015})},\ \Eprint {http://arxiv.org/abs/1405.2925}
  {arXiv:1405.2925 [hep-ph]} \BibitemShut {NoStop}%
\bibitem [{\citenamefont {Arvanitaki}\ \emph {et~al.}(2016)\citenamefont
  {Arvanitaki}, \citenamefont {Dimopoulos},\ and\ \citenamefont
  {Van~Tilburg}}]{Arvanitaki:2015iga}%
  \BibitemOpen
  \bibfield  {author} {\bibinfo {author} {\bibfnamefont {A.}~\bibnamefont
  {Arvanitaki}}, \bibinfo {author} {\bibfnamefont {S.}~\bibnamefont
  {Dimopoulos}}, \ and\ \bibinfo {author} {\bibfnamefont {K.}~\bibnamefont
  {Van~Tilburg}},\ }\href {\doibase 10.1103/PhysRevLett.116.031102} {\bibfield
  {journal} {\bibinfo  {journal} {Phys. Rev. Lett.}\ }\textbf {\bibinfo
  {volume} {116}},\ \bibinfo {pages} {031102} (\bibinfo {year} {2016})},\
  \Eprint {http://arxiv.org/abs/1508.01798} {arXiv:1508.01798 [hep-ph]}
  \BibitemShut {NoStop}%
\bibitem [{\citenamefont {Branca}\ \emph {et~al.}(2017)\citenamefont {Branca}
  \emph {et~al.}}]{Branca:2016rez}%
  \BibitemOpen
  \bibfield  {author} {\bibinfo {author} {\bibfnamefont {A.}~\bibnamefont
  {Branca}} \emph {et~al.},\ }\href {\doibase 10.1103/PhysRevLett.118.021302}
  {\bibfield  {journal} {\bibinfo  {journal} {Phys. Rev. Lett.}\ }\textbf
  {\bibinfo {volume} {118}},\ \bibinfo {pages} {021302} (\bibinfo {year}
  {2017})},\ \Eprint {http://arxiv.org/abs/1607.07327} {arXiv:1607.07327
  [hep-ex]} \BibitemShut {NoStop}%
\bibitem [{\citenamefont {Graham}\ \emph {et~al.}(2016)\citenamefont {Graham},
  \citenamefont {Kaplan}, \citenamefont {Mardon}, \citenamefont {Rajendran},\
  and\ \citenamefont {Terrano}}]{Graham:2015ifn}%
  \BibitemOpen
  \bibfield  {author} {\bibinfo {author} {\bibfnamefont {P.~W.}\ \bibnamefont
  {Graham}}, \bibinfo {author} {\bibfnamefont {D.~E.}\ \bibnamefont {Kaplan}},
  \bibinfo {author} {\bibfnamefont {J.}~\bibnamefont {Mardon}}, \bibinfo
  {author} {\bibfnamefont {S.}~\bibnamefont {Rajendran}}, \ and\ \bibinfo
  {author} {\bibfnamefont {W.~A.}\ \bibnamefont {Terrano}},\ }\href {\doibase
  10.1103/PhysRevD.93.075029} {\bibfield  {journal} {\bibinfo  {journal} {Phys.
  Rev.}\ }\textbf {\bibinfo {volume} {D93}},\ \bibinfo {pages} {075029}
  (\bibinfo {year} {2016})},\ \Eprint {http://arxiv.org/abs/1512.06165}
  {arXiv:1512.06165 [hep-ph]} \BibitemShut {NoStop}%
\bibitem [{\citenamefont {Blas}\ \emph {et~al.}(2017)\citenamefont {Blas},
  \citenamefont {Nacir},\ and\ \citenamefont {Sibiryakov}}]{Blas:2016ddr}%
  \BibitemOpen
  \bibfield  {author} {\bibinfo {author} {\bibfnamefont {D.}~\bibnamefont
  {Blas}}, \bibinfo {author} {\bibfnamefont {D.~L.}\ \bibnamefont {Nacir}}, \
  and\ \bibinfo {author} {\bibfnamefont {S.}~\bibnamefont {Sibiryakov}},\
  }\href {\doibase 10.1103/PhysRevLett.118.261102} {\bibfield  {journal}
  {\bibinfo  {journal} {Phys. Rev. Lett.}\ }\textbf {\bibinfo {volume} {118}},\
  \bibinfo {pages} {261102} (\bibinfo {year} {2017})},\ \Eprint
  {http://arxiv.org/abs/1612.06789} {arXiv:1612.06789 [hep-ph]} \BibitemShut
  {NoStop}%
\bibitem [{\citenamefont {Arvanitaki}\ \emph {et~al.}(2018)\citenamefont
  {Arvanitaki}, \citenamefont {Graham}, \citenamefont {Hogan}, \citenamefont
  {Rajendran},\ and\ \citenamefont {Van~Tilburg}}]{Arvanitaki:2016fyj}%
  \BibitemOpen
  \bibfield  {author} {\bibinfo {author} {\bibfnamefont {A.}~\bibnamefont
  {Arvanitaki}}, \bibinfo {author} {\bibfnamefont {P.~W.}\ \bibnamefont
  {Graham}}, \bibinfo {author} {\bibfnamefont {J.~M.}\ \bibnamefont {Hogan}},
  \bibinfo {author} {\bibfnamefont {S.}~\bibnamefont {Rajendran}}, \ and\
  \bibinfo {author} {\bibfnamefont {K.}~\bibnamefont {Van~Tilburg}},\ }\href
  {\doibase 10.1103/PhysRevD.97.075020} {\bibfield  {journal} {\bibinfo
  {journal} {Phys. Rev.}\ }\textbf {\bibinfo {volume} {D97}},\ \bibinfo {pages}
  {075020} (\bibinfo {year} {2018})},\ \Eprint
  {http://arxiv.org/abs/1606.04541} {arXiv:1606.04541 [hep-ph]} \BibitemShut
  {NoStop}%
\bibitem [{\citenamefont {Geraci}\ \emph {et~al.}(2018)\citenamefont {Geraci},
  \citenamefont {Bradley}, \citenamefont {Gao}, \citenamefont {Weinstein},\
  and\ \citenamefont {Derevianko}}]{Geraci:2018fax}%
  \BibitemOpen
  \bibfield  {author} {\bibinfo {author} {\bibfnamefont {A.~A.}\ \bibnamefont
  {Geraci}}, \bibinfo {author} {\bibfnamefont {C.}~\bibnamefont {Bradley}},
  \bibinfo {author} {\bibfnamefont {D.}~\bibnamefont {Gao}}, \bibinfo {author}
  {\bibfnamefont {J.}~\bibnamefont {Weinstein}}, \ and\ \bibinfo {author}
  {\bibfnamefont {A.}~\bibnamefont {Derevianko}},\ }\href@noop {} {\  (\bibinfo
  {year} {2018})},\ \Eprint {http://arxiv.org/abs/1808.00540} {arXiv:1808.00540
  [astro-ph.IM]} \BibitemShut {NoStop}%
\bibitem [{\citenamefont {Hees}\ \emph {et~al.}(2016)\citenamefont {Hees},
  \citenamefont {Gu\'ena}, \citenamefont {Abgrall}, \citenamefont {Bize},\ and\
  \citenamefont {Wolf}}]{Hees:2016gop}%
  \BibitemOpen
  \bibfield  {author} {\bibinfo {author} {\bibfnamefont {A.}~\bibnamefont
  {Hees}}, \bibinfo {author} {\bibfnamefont {J.}~\bibnamefont {Gu\'ena}},
  \bibinfo {author} {\bibfnamefont {M.}~\bibnamefont {Abgrall}}, \bibinfo
  {author} {\bibfnamefont {S.}~\bibnamefont {Bize}}, \ and\ \bibinfo {author}
  {\bibfnamefont {P.}~\bibnamefont {Wolf}},\ }\href {\doibase
  10.1103/PhysRevLett.117.061301} {\bibfield  {journal} {\bibinfo  {journal}
  {Phys. Rev. Lett.}\ }\textbf {\bibinfo {volume} {117}},\ \bibinfo {pages}
  {061301} (\bibinfo {year} {2016})},\ \Eprint
  {http://arxiv.org/abs/1604.08514} {arXiv:1604.08514 [gr-qc]} \BibitemShut
  {NoStop}%
\bibitem [{\citenamefont {Stadnik}\ and\ \citenamefont
  {Flambaum}(2016)}]{Stadnik:2015xbn}%
  \BibitemOpen
  \bibfield  {author} {\bibinfo {author} {\bibfnamefont {Y.~V.}\ \bibnamefont
  {Stadnik}}\ and\ \bibinfo {author} {\bibfnamefont {V.~V.}\ \bibnamefont
  {Flambaum}},\ }\href {\doibase 10.1103/PhysRevA.93.063630} {\bibfield
  {journal} {\bibinfo  {journal} {Phys. Rev.}\ }\textbf {\bibinfo {volume}
  {A93}},\ \bibinfo {pages} {063630} (\bibinfo {year} {2016})},\ \Eprint
  {http://arxiv.org/abs/1511.00447} {arXiv:1511.00447 [physics.atom-ph]}
  \BibitemShut {NoStop}%
\bibitem [{\citenamefont {Aoki}\ and\ \citenamefont
  {Soda}(2016)}]{Aoki:2016kwl}%
  \BibitemOpen
  \bibfield  {author} {\bibinfo {author} {\bibfnamefont {A.}~\bibnamefont
  {Aoki}}\ and\ \bibinfo {author} {\bibfnamefont {J.}~\bibnamefont {Soda}},\
  }\href {\doibase 10.1142/S0218271817500638} {\bibfield  {journal} {\bibinfo
  {journal} {Int. J. Mod. Phys.}\ }\textbf {\bibinfo {volume} {D26}},\ \bibinfo
  {pages} {1750063} (\bibinfo {year} {2016})},\ \Eprint
  {http://arxiv.org/abs/1608.05933} {arXiv:1608.05933 [astro-ph.CO]}
  \BibitemShut {NoStop}%
\bibitem [{\citenamefont {Morisaki}\ and\ \citenamefont
  {Suyama}(2019)}]{Morisaki:2018htj}%
  \BibitemOpen
  \bibfield  {author} {\bibinfo {author} {\bibfnamefont {S.}~\bibnamefont
  {Morisaki}}\ and\ \bibinfo {author} {\bibfnamefont {T.}~\bibnamefont
  {Suyama}},\ }\href {\doibase 10.1103/PhysRevD.100.123512} {\bibfield
  {journal} {\bibinfo  {journal} {Phys. Rev. D}\ }\textbf {\bibinfo {volume}
  {100}},\ \bibinfo {pages} {123512} (\bibinfo {year} {2019})},\ \Eprint
  {http://arxiv.org/abs/1811.05003} {arXiv:1811.05003 [hep-ph]} \BibitemShut
  {NoStop}%
\bibitem [{\citenamefont {DeRocco}\ and\ \citenamefont
  {Hook}(2018)}]{DeRocco:2018jwe}%
  \BibitemOpen
  \bibfield  {author} {\bibinfo {author} {\bibfnamefont {W.}~\bibnamefont
  {DeRocco}}\ and\ \bibinfo {author} {\bibfnamefont {A.}~\bibnamefont {Hook}},\
  }\href {\doibase 10.1103/PhysRevD.98.035021} {\bibfield  {journal} {\bibinfo
  {journal} {Phys. Rev. D}\ }\textbf {\bibinfo {volume} {98}},\ \bibinfo
  {pages} {035021} (\bibinfo {year} {2018})},\ \Eprint
  {http://arxiv.org/abs/1802.07273} {arXiv:1802.07273 [hep-ph]} \BibitemShut
  {NoStop}%
\bibitem [{\citenamefont {Obata}\ \emph {et~al.}(2018)\citenamefont {Obata},
  \citenamefont {Fujita},\ and\ \citenamefont {Michimura}}]{Obata:2018vvr}%
  \BibitemOpen
  \bibfield  {author} {\bibinfo {author} {\bibfnamefont {I.}~\bibnamefont
  {Obata}}, \bibinfo {author} {\bibfnamefont {T.}~\bibnamefont {Fujita}}, \
  and\ \bibinfo {author} {\bibfnamefont {Y.}~\bibnamefont {Michimura}},\ }\href
  {\doibase 10.1103/PhysRevLett.121.161301} {\bibfield  {journal} {\bibinfo
  {journal} {Phys. Rev. Lett.}\ }\textbf {\bibinfo {volume} {121}},\ \bibinfo
  {pages} {161301} (\bibinfo {year} {2018})},\ \Eprint
  {http://arxiv.org/abs/1805.11753} {arXiv:1805.11753 [astro-ph.CO]}
  \BibitemShut {NoStop}%
\bibitem [{\citenamefont {Liu}\ \emph {et~al.}(2019)\citenamefont {Liu},
  \citenamefont {Elwood}, \citenamefont {Evans},\ and\ \citenamefont
  {Thaler}}]{Liu:2018icu}%
  \BibitemOpen
  \bibfield  {author} {\bibinfo {author} {\bibfnamefont {H.}~\bibnamefont
  {Liu}}, \bibinfo {author} {\bibfnamefont {B.~D.}\ \bibnamefont {Elwood}},
  \bibinfo {author} {\bibfnamefont {M.}~\bibnamefont {Evans}}, \ and\ \bibinfo
  {author} {\bibfnamefont {J.}~\bibnamefont {Thaler}},\ }\href {\doibase
  10.1103/PhysRevD.100.023548} {\bibfield  {journal} {\bibinfo  {journal}
  {Phys. Rev. D}\ }\textbf {\bibinfo {volume} {100}},\ \bibinfo {pages}
  {023548} (\bibinfo {year} {2019})},\ \Eprint
  {http://arxiv.org/abs/1809.01656} {arXiv:1809.01656 [hep-ph]} \BibitemShut
  {NoStop}%
\bibitem [{\citenamefont {Nagano}\ \emph {et~al.}(2019)\citenamefont {Nagano},
  \citenamefont {Fujita}, \citenamefont {Michimura},\ and\ \citenamefont
  {Obata}}]{Nagano:2019rbw}%
  \BibitemOpen
  \bibfield  {author} {\bibinfo {author} {\bibfnamefont {K.}~\bibnamefont
  {Nagano}}, \bibinfo {author} {\bibfnamefont {T.}~\bibnamefont {Fujita}},
  \bibinfo {author} {\bibfnamefont {Y.}~\bibnamefont {Michimura}}, \ and\
  \bibinfo {author} {\bibfnamefont {I.}~\bibnamefont {Obata}},\ }\href
  {\doibase 10.1103/PhysRevLett.123.111301} {\bibfield  {journal} {\bibinfo
  {journal} {Phys. Rev. Lett.}\ }\textbf {\bibinfo {volume} {123}},\ \bibinfo
  {pages} {111301} (\bibinfo {year} {2019})},\ \Eprint
  {http://arxiv.org/abs/1903.02017} {arXiv:1903.02017 [hep-ph]} \BibitemShut
  {NoStop}%
\bibitem [{\citenamefont {Martynov}\ and\ \citenamefont
  {Miao}(2020)}]{Martynov:2019azm}%
  \BibitemOpen
  \bibfield  {author} {\bibinfo {author} {\bibfnamefont {D.}~\bibnamefont
  {Martynov}}\ and\ \bibinfo {author} {\bibfnamefont {H.}~\bibnamefont
  {Miao}},\ }\href {\doibase 10.1103/PhysRevD.101.095034} {\bibfield  {journal}
  {\bibinfo  {journal} {Phys. Rev. D}\ }\textbf {\bibinfo {volume} {101}},\
  \bibinfo {pages} {095034} (\bibinfo {year} {2020})},\ \Eprint
  {http://arxiv.org/abs/1911.00429} {arXiv:1911.00429 [physics.ins-det]}
  \BibitemShut {NoStop}%
\bibitem [{\citenamefont {Michimura}\ \emph {et~al.}(2020)\citenamefont
  {Michimura}, \citenamefont {Fujita}, \citenamefont {Morisaki}, \citenamefont
  {Nakatsuka},\ and\ \citenamefont {Obata}}]{Michimura:2020vxn}%
  \BibitemOpen
  \bibfield  {author} {\bibinfo {author} {\bibfnamefont {Y.}~\bibnamefont
  {Michimura}}, \bibinfo {author} {\bibfnamefont {T.}~\bibnamefont {Fujita}},
  \bibinfo {author} {\bibfnamefont {S.}~\bibnamefont {Morisaki}}, \bibinfo
  {author} {\bibfnamefont {H.}~\bibnamefont {Nakatsuka}}, \ and\ \bibinfo
  {author} {\bibfnamefont {I.}~\bibnamefont {Obata}},\ }\href {\doibase
  10.1103/PhysRevD.102.102001} {\bibfield  {journal} {\bibinfo  {journal}
  {Phys. Rev. D}\ }\textbf {\bibinfo {volume} {102}},\ \bibinfo {pages}
  {102001} (\bibinfo {year} {2020})},\ \Eprint
  {http://arxiv.org/abs/2008.02482} {arXiv:2008.02482 [hep-ph]} \BibitemShut
  {NoStop}%
\bibitem [{\citenamefont {Pierce}\ \emph {et~al.}(2018)\citenamefont {Pierce},
  \citenamefont {Riles},\ and\ \citenamefont {Zhao}}]{Pierce:2018xmy}%
  \BibitemOpen
  \bibfield  {author} {\bibinfo {author} {\bibfnamefont {A.}~\bibnamefont
  {Pierce}}, \bibinfo {author} {\bibfnamefont {K.}~\bibnamefont {Riles}}, \
  and\ \bibinfo {author} {\bibfnamefont {Y.}~\bibnamefont {Zhao}},\ }\href
  {\doibase 10.1103/PhysRevLett.121.061102} {\bibfield  {journal} {\bibinfo
  {journal} {Phys. Rev. Lett.}\ }\textbf {\bibinfo {volume} {121}},\ \bibinfo
  {pages} {061102} (\bibinfo {year} {2018})},\ \Eprint
  {http://arxiv.org/abs/1801.10161} {arXiv:1801.10161 [hep-ph]} \BibitemShut
  {NoStop}%
\bibitem [{\citenamefont {Guo}\ \emph {et~al.}(2019)\citenamefont {Guo},
  \citenamefont {Riles}, \citenamefont {Yang},\ and\ \citenamefont
  {Zhao}}]{Guo:2019ker}%
  \BibitemOpen
  \bibfield  {author} {\bibinfo {author} {\bibfnamefont {H.-K.}\ \bibnamefont
  {Guo}}, \bibinfo {author} {\bibfnamefont {K.}~\bibnamefont {Riles}}, \bibinfo
  {author} {\bibfnamefont {F.-W.}\ \bibnamefont {Yang}}, \ and\ \bibinfo
  {author} {\bibfnamefont {Y.}~\bibnamefont {Zhao}},\ }\href {\doibase
  10.1038/s42005-019-0255-0} {\bibfield  {journal} {\bibinfo  {journal}
  {Commun. Phys.}\ }\textbf {\bibinfo {volume} {2}},\ \bibinfo {pages} {155}
  (\bibinfo {year} {2019})},\ \Eprint {http://arxiv.org/abs/1905.04316}
  {arXiv:1905.04316 [hep-ph]} \BibitemShut {NoStop}%
\bibitem [{\citenamefont {Grote}\ and\ \citenamefont
  {Stadnik}(2019)}]{Grote:2019uvn}%
  \BibitemOpen
  \bibfield  {author} {\bibinfo {author} {\bibfnamefont {H.}~\bibnamefont
  {Grote}}\ and\ \bibinfo {author} {\bibfnamefont {Y.}~\bibnamefont
  {Stadnik}},\ }\href {\doibase 10.1103/PhysRevResearch.1.033187} {\bibfield
  {journal} {\bibinfo  {journal} {Phys. Rev. Res.}\ }\textbf {\bibinfo {volume}
  {1}},\ \bibinfo {pages} {033187} (\bibinfo {year} {2019})},\ \Eprint
  {http://arxiv.org/abs/1906.06193} {arXiv:1906.06193 [astro-ph.IM]}
  \BibitemShut {NoStop}%
\bibitem [{\citenamefont {Calder\'on~Bustillo}\ \emph
  {et~al.}(2020)\citenamefont {Calder\'on~Bustillo}, \citenamefont
  {Sanchis-Gual}, \citenamefont {Torres-Forn\'e}, \citenamefont {Font},
  \citenamefont {Vajpeyi}, \citenamefont {Smith}, \citenamefont {Herdeiro},
  \citenamefont {Radu},\ and\ \citenamefont
  {Leong}}]{CalderonBustillo:2020srq}%
  \BibitemOpen
  \bibfield  {author} {\bibinfo {author} {\bibfnamefont {J.}~\bibnamefont
  {Calder\'on~Bustillo}}, \bibinfo {author} {\bibfnamefont {N.}~\bibnamefont
  {Sanchis-Gual}}, \bibinfo {author} {\bibfnamefont {A.}~\bibnamefont
  {Torres-Forn\'e}}, \bibinfo {author} {\bibfnamefont {J.~A.}\ \bibnamefont
  {Font}}, \bibinfo {author} {\bibfnamefont {A.}~\bibnamefont {Vajpeyi}},
  \bibinfo {author} {\bibfnamefont {R.}~\bibnamefont {Smith}}, \bibinfo
  {author} {\bibfnamefont {C.}~\bibnamefont {Herdeiro}}, \bibinfo {author}
  {\bibfnamefont {E.}~\bibnamefont {Radu}}, \ and\ \bibinfo {author}
  {\bibfnamefont {S.~H.}\ \bibnamefont {Leong}},\ }\href@noop {} {\  (\bibinfo
  {year} {2020})},\ \Eprint {http://arxiv.org/abs/2009.05376} {arXiv:2009.05376
  [gr-qc]} \BibitemShut {NoStop}%
\bibitem [{\citenamefont {Fujita}\ \emph {et~al.}(2019)\citenamefont {Fujita},
  \citenamefont {Tazaki},\ and\ \citenamefont {Toma}}]{Fujita:2018zaj}%
  \BibitemOpen
  \bibfield  {author} {\bibinfo {author} {\bibfnamefont {T.}~\bibnamefont
  {Fujita}}, \bibinfo {author} {\bibfnamefont {R.}~\bibnamefont {Tazaki}}, \
  and\ \bibinfo {author} {\bibfnamefont {K.}~\bibnamefont {Toma}},\ }\href
  {\doibase 10.1103/PhysRevLett.122.191101} {\bibfield  {journal} {\bibinfo
  {journal} {Phys. Rev. Lett.}\ }\textbf {\bibinfo {volume} {122}},\ \bibinfo
  {pages} {191101} (\bibinfo {year} {2019})},\ \Eprint
  {http://arxiv.org/abs/1811.03525} {arXiv:1811.03525 [astro-ph.CO]}
  \BibitemShut {NoStop}%
\bibitem [{\citenamefont {Caputo}\ \emph {et~al.}(2019)\citenamefont {Caputo},
  \citenamefont {Sberna}, \citenamefont {Frias}, \citenamefont {Blas},
  \citenamefont {Pani}, \citenamefont {Shao},\ and\ \citenamefont
  {Yan}}]{Caputo:2019tms}%
  \BibitemOpen
  \bibfield  {author} {\bibinfo {author} {\bibfnamefont {A.}~\bibnamefont
  {Caputo}}, \bibinfo {author} {\bibfnamefont {L.}~\bibnamefont {Sberna}},
  \bibinfo {author} {\bibfnamefont {M.}~\bibnamefont {Frias}}, \bibinfo
  {author} {\bibfnamefont {D.}~\bibnamefont {Blas}}, \bibinfo {author}
  {\bibfnamefont {P.}~\bibnamefont {Pani}}, \bibinfo {author} {\bibfnamefont
  {L.}~\bibnamefont {Shao}}, \ and\ \bibinfo {author} {\bibfnamefont
  {W.}~\bibnamefont {Yan}},\ }\href {\doibase 10.1103/PhysRevD.100.063515}
  {\bibfield  {journal} {\bibinfo  {journal} {Phys. Rev. D}\ }\textbf {\bibinfo
  {volume} {100}},\ \bibinfo {pages} {063515} (\bibinfo {year} {2019})},\
  \Eprint {http://arxiv.org/abs/1902.02695} {arXiv:1902.02695 [astro-ph.CO]}
  \BibitemShut {NoStop}%
\bibitem [{\citenamefont {Fedderke}\ \emph {et~al.}(2019)\citenamefont
  {Fedderke}, \citenamefont {Graham},\ and\ \citenamefont
  {Rajendran}}]{Fedderke:2019ajk}%
  \BibitemOpen
  \bibfield  {author} {\bibinfo {author} {\bibfnamefont {M.~A.}\ \bibnamefont
  {Fedderke}}, \bibinfo {author} {\bibfnamefont {P.~W.}\ \bibnamefont
  {Graham}}, \ and\ \bibinfo {author} {\bibfnamefont {S.}~\bibnamefont
  {Rajendran}},\ }\href {\doibase 10.1103/PhysRevD.100.015040} {\bibfield
  {journal} {\bibinfo  {journal} {Phys. Rev. D}\ }\textbf {\bibinfo {volume}
  {100}},\ \bibinfo {pages} {015040} (\bibinfo {year} {2019})},\ \Eprint
  {http://arxiv.org/abs/1903.02666} {arXiv:1903.02666 [astro-ph.CO]}
  \BibitemShut {NoStop}%
\bibitem [{\citenamefont {Schlamminger}\ \emph {et~al.}(2008)\citenamefont
  {Schlamminger}, \citenamefont {Choi}, \citenamefont {Wagner}, \citenamefont
  {Gundlach},\ and\ \citenamefont {Adelberger}}]{Schlamminger:2007ht}%
  \BibitemOpen
  \bibfield  {author} {\bibinfo {author} {\bibfnamefont {S.}~\bibnamefont
  {Schlamminger}}, \bibinfo {author} {\bibfnamefont {K.-Y.}\ \bibnamefont
  {Choi}}, \bibinfo {author} {\bibfnamefont {T.}~\bibnamefont {Wagner}},
  \bibinfo {author} {\bibfnamefont {J.}~\bibnamefont {Gundlach}}, \ and\
  \bibinfo {author} {\bibfnamefont {E.}~\bibnamefont {Adelberger}},\ }\href
  {\doibase 10.1103/PhysRevLett.100.041101} {\bibfield  {journal} {\bibinfo
  {journal} {Phys. Rev. Lett.}\ }\textbf {\bibinfo {volume} {100}},\ \bibinfo
  {pages} {041101} (\bibinfo {year} {2008})},\ \Eprint
  {http://arxiv.org/abs/0712.0607} {arXiv:0712.0607 [gr-qc]} \BibitemShut
  {NoStop}%
\bibitem [{\citenamefont {Wagner}\ \emph {et~al.}(2012)\citenamefont {Wagner},
  \citenamefont {Schlamminger}, \citenamefont {Gundlach},\ and\ \citenamefont
  {Adelberger}}]{Wagner:2012ui}%
  \BibitemOpen
  \bibfield  {author} {\bibinfo {author} {\bibfnamefont {T.}~\bibnamefont
  {Wagner}}, \bibinfo {author} {\bibfnamefont {S.}~\bibnamefont
  {Schlamminger}}, \bibinfo {author} {\bibfnamefont {J.}~\bibnamefont
  {Gundlach}}, \ and\ \bibinfo {author} {\bibfnamefont {E.}~\bibnamefont
  {Adelberger}},\ }\href {\doibase 10.1088/0264-9381/29/18/184002} {\bibfield
  {journal} {\bibinfo  {journal} {Class. Quant. Grav.}\ }\textbf {\bibinfo
  {volume} {29}},\ \bibinfo {pages} {184002} (\bibinfo {year} {2012})},\
  \Eprint {http://arxiv.org/abs/1207.2442} {arXiv:1207.2442 [gr-qc]}
  \BibitemShut {NoStop}%
\bibitem [{\citenamefont {Touboul}\ \emph {et~al.}(2017)\citenamefont {Touboul}
  \emph {et~al.}}]{Touboul:2017grn}%
  \BibitemOpen
  \bibfield  {author} {\bibinfo {author} {\bibfnamefont {P.}~\bibnamefont
  {Touboul}} \emph {et~al.},\ }\href {\doibase 10.1103/PhysRevLett.119.231101}
  {\bibfield  {journal} {\bibinfo  {journal} {Phys. Rev. Lett.}\ }\textbf
  {\bibinfo {volume} {119}},\ \bibinfo {pages} {231101} (\bibinfo {year}
  {2017})},\ \Eprint {http://arxiv.org/abs/1712.01176} {arXiv:1712.01176
  [astro-ph.IM]} \BibitemShut {NoStop}%
\bibitem [{\citenamefont {Bergé}\ \emph {et~al.}(2018)\citenamefont {Bergé},
  \citenamefont {Brax}, \citenamefont {Métris}, \citenamefont
  {Pernot-Borràs}, \citenamefont {Touboul},\ and\ \citenamefont
  {Uzan}}]{Berge:2017ovy}%
  \BibitemOpen
  \bibfield  {author} {\bibinfo {author} {\bibfnamefont {J.}~\bibnamefont
  {Bergé}}, \bibinfo {author} {\bibfnamefont {P.}~\bibnamefont {Brax}},
  \bibinfo {author} {\bibfnamefont {G.}~\bibnamefont {Métris}}, \bibinfo
  {author} {\bibfnamefont {M.}~\bibnamefont {Pernot-Borràs}}, \bibinfo
  {author} {\bibfnamefont {P.}~\bibnamefont {Touboul}}, \ and\ \bibinfo
  {author} {\bibfnamefont {J.-P.}\ \bibnamefont {Uzan}},\ }\href {\doibase
  10.1103/PhysRevLett.120.141101} {\bibfield  {journal} {\bibinfo  {journal}
  {Phys. Rev. Lett.}\ }\textbf {\bibinfo {volume} {120}},\ \bibinfo {pages}
  {141101} (\bibinfo {year} {2018})},\ \Eprint
  {http://arxiv.org/abs/1712.00483} {arXiv:1712.00483 [gr-qc]} \BibitemShut
  {NoStop}%
\bibitem [{\citenamefont {Harry}(2010)}]{Harry:2010zz}%
  \BibitemOpen
  \bibfield  {author} {\bibinfo {author} {\bibfnamefont {G.~M.}\ \bibnamefont
  {Harry}} (\bibinfo {collaboration} {LIGO Scientific}),\ }\href {\doibase
  10.1088/0264-9381/27/8/084006} {\bibfield  {journal} {\bibinfo  {journal}
  {Class. Quant. Grav.}\ }\textbf {\bibinfo {volume} {27}},\ \bibinfo {pages}
  {084006} (\bibinfo {year} {2010})}\BibitemShut {NoStop}%
\bibitem [{\citenamefont {Miller}\ \emph {et~al.}(2020)\citenamefont {Miller}
  \emph {et~al.}}]{Miller:2020vsl}%
  \BibitemOpen
  \bibfield  {author} {\bibinfo {author} {\bibfnamefont {A.~L.}\ \bibnamefont
  {Miller}} \emph {et~al.},\ }\href@noop {} {\  (\bibinfo {year} {2020})},\
  \Eprint {http://arxiv.org/abs/2010.01925} {arXiv:2010.01925 [astro-ph.IM]}
  \BibitemShut {NoStop}%
\bibitem [{\citenamefont {Hild}\ \emph {et~al.}(2011)\citenamefont {Hild} \emph
  {et~al.}}]{Hild:2010id}%
  \BibitemOpen
  \bibfield  {author} {\bibinfo {author} {\bibfnamefont {S.}~\bibnamefont
  {Hild}} \emph {et~al.},\ }\href {\doibase 10.1088/0264-9381/28/9/094013}
  {\bibfield  {journal} {\bibinfo  {journal} {Class. Quant. Grav.}\ }\textbf
  {\bibinfo {volume} {28}},\ \bibinfo {pages} {094013} (\bibinfo {year}
  {2011})},\ \Eprint {http://arxiv.org/abs/1012.0908} {arXiv:1012.0908 [gr-qc]}
  \BibitemShut {NoStop}%
\bibitem [{\citenamefont {Abbott}\ \emph {et~al.}(2017)\citenamefont {Abbott}
  \emph {et~al.}}]{Evans:2016mbw}%
  \BibitemOpen
  \bibfield  {author} {\bibinfo {author} {\bibfnamefont {B.~P.}\ \bibnamefont
  {Abbott}} \emph {et~al.} (\bibinfo {collaboration} {LIGO Scientific}),\
  }\href {\doibase 10.1088/1361-6382/aa51f4} {\bibfield  {journal} {\bibinfo
  {journal} {Class. Quant. Grav.}\ }\textbf {\bibinfo {volume} {34}},\ \bibinfo
  {pages} {044001} (\bibinfo {year} {2017})},\ \Eprint
  {http://arxiv.org/abs/1607.08697} {arXiv:1607.08697 [astro-ph.IM]}
  \BibitemShut {NoStop}%
\bibitem [{\citenamefont {Kawamura}\ \emph {et~al.}(2006)\citenamefont
  {Kawamura} \emph {et~al.}}]{Kawamura:2006up}%
  \BibitemOpen
  \bibfield  {author} {\bibinfo {author} {\bibfnamefont {S.}~\bibnamefont
  {Kawamura}} \emph {et~al.},\ }\href {\doibase 10.1088/0264-9381/23/8/S17}
  {\bibfield  {journal} {\bibinfo  {journal} {Class. Quant. Grav.}\ }\textbf
  {\bibinfo {volume} {23}},\ \bibinfo {pages} {S125} (\bibinfo {year}
  {2006})}\BibitemShut {NoStop}%
\bibitem [{\citenamefont {Amaro-Seoane}\ \emph {et~al.}(2017)\citenamefont
  {Amaro-Seoane} \emph {et~al.}}]{Audley:2017drz}%
  \BibitemOpen
  \bibfield  {author} {\bibinfo {author} {\bibfnamefont {P.}~\bibnamefont
  {Amaro-Seoane}} \emph {et~al.} (\bibinfo {collaboration} {LISA}),\
  }\href@noop {} {\  (\bibinfo {year} {2017})},\ \Eprint
  {http://arxiv.org/abs/1702.00786} {arXiv:1702.00786 [astro-ph.IM]}
  \BibitemShut {NoStop}%
\bibitem [{\citenamefont {Smith}\ \emph {et~al.}(2007)\citenamefont {Smith}
  \emph {et~al.}}]{Smith:2006ym}%
  \BibitemOpen
  \bibfield  {author} {\bibinfo {author} {\bibfnamefont {M.~C.}\ \bibnamefont
  {Smith}} \emph {et~al.},\ }\href {\doibase 10.1111/j.1365-2966.2007.11964.x}
  {\bibfield  {journal} {\bibinfo  {journal} {Mon. Not. Roy. Astron. Soc.}\
  }\textbf {\bibinfo {volume} {379}},\ \bibinfo {pages} {755} (\bibinfo {year}
  {2007})},\ \Eprint {http://arxiv.org/abs/astro-ph/0611671}
  {arXiv:astro-ph/0611671} \BibitemShut {NoStop}%
\bibitem [{\citenamefont {Allen}\ and\ \citenamefont
  {Romano}(1999)}]{Allen:1997ad}%
  \BibitemOpen
  \bibfield  {author} {\bibinfo {author} {\bibfnamefont {B.}~\bibnamefont
  {Allen}}\ and\ \bibinfo {author} {\bibfnamefont {J.~D.}\ \bibnamefont
  {Romano}},\ }\href {\doibase 10.1103/PhysRevD.59.102001} {\bibfield
  {journal} {\bibinfo  {journal} {Phys. Rev. D}\ }\textbf {\bibinfo {volume}
  {59}},\ \bibinfo {pages} {102001} (\bibinfo {year} {1999})},\ \Eprint
  {http://arxiv.org/abs/gr-qc/9710117} {arXiv:gr-qc/9710117} \BibitemShut
  {NoStop}%
\bibitem [{\citenamefont {Larson}\ \emph {et~al.}(2000)\citenamefont {Larson},
  \citenamefont {Hiscock},\ and\ \citenamefont {Hellings}}]{Larson:1999we}%
  \BibitemOpen
  \bibfield  {author} {\bibinfo {author} {\bibfnamefont {S.~L.}\ \bibnamefont
  {Larson}}, \bibinfo {author} {\bibfnamefont {W.~A.}\ \bibnamefont {Hiscock}},
  \ and\ \bibinfo {author} {\bibfnamefont {R.~W.}\ \bibnamefont {Hellings}},\
  }\href {\doibase 10.1103/PhysRevD.62.062001} {\bibfield  {journal} {\bibinfo
  {journal} {Phys. Rev. D}\ }\textbf {\bibinfo {volume} {62}},\ \bibinfo
  {pages} {062001} (\bibinfo {year} {2000})},\ \Eprint
  {http://arxiv.org/abs/gr-qc/9909080} {arXiv:gr-qc/9909080} \BibitemShut
  {NoStop}%
\bibitem [{ali(2018)}]{aligosensitivity}%
  \BibitemOpen
  \href@noop {} {\emph {\bibinfo {title} {{Updated Advanced LIGO sensitivity
  design curve}}}} (\bibinfo {year} {2018}),\ \bibinfo {note} {{{LIGO Document
  T1800044-v4 \url{https://dcc.ligo.org/LIGO-T1800044/public}}}}\BibitemShut
  {NoStop}%
\bibitem [{\citenamefont {Essick}\ \emph {et~al.}(2017)\citenamefont {Essick},
  \citenamefont {Vitale},\ and\ \citenamefont {Evans}}]{Essick:2017wyl}%
  \BibitemOpen
  \bibfield  {author} {\bibinfo {author} {\bibfnamefont {R.}~\bibnamefont
  {Essick}}, \bibinfo {author} {\bibfnamefont {S.}~\bibnamefont {Vitale}}, \
  and\ \bibinfo {author} {\bibfnamefont {M.}~\bibnamefont {Evans}},\ }\href
  {\doibase 10.1103/PhysRevD.96.084004} {\bibfield  {journal} {\bibinfo
  {journal} {Phys. Rev. D}\ }\textbf {\bibinfo {volume} {96}},\ \bibinfo
  {pages} {084004} (\bibinfo {year} {2017})},\ \Eprint
  {http://arxiv.org/abs/1708.06843} {arXiv:1708.06843 [gr-qc]} \BibitemShut
  {NoStop}%
\bibitem [{\citenamefont {Kawamura}\ \emph {et~al.}(2020)\citenamefont
  {Kawamura} \emph {et~al.}}]{Kawamura:2020pcg}%
  \BibitemOpen
  \bibfield  {author} {\bibinfo {author} {\bibfnamefont {S.}~\bibnamefont
  {Kawamura}} \emph {et~al.},\ }\href@noop {} {\  (\bibinfo {year} {2020})},\
  \Eprint {http://arxiv.org/abs/2006.13545} {arXiv:2006.13545 [gr-qc]}
  \BibitemShut {NoStop}%
\bibitem [{\citenamefont {Fayet}(2018)}]{Fayet:2017pdp}%
  \BibitemOpen
  \bibfield  {author} {\bibinfo {author} {\bibfnamefont {P.}~\bibnamefont
  {Fayet}},\ }\href {\doibase 10.1103/PhysRevD.97.055039} {\bibfield  {journal}
  {\bibinfo  {journal} {Phys. Rev. D}\ }\textbf {\bibinfo {volume} {97}},\
  \bibinfo {pages} {055039} (\bibinfo {year} {2018})},\ \Eprint
  {http://arxiv.org/abs/1712.00856} {arXiv:1712.00856 [hep-ph]} \BibitemShut
  {NoStop}%
\end{thebibliography}%

\end{document}